\documentclass[twocolumn,showpacs,preprintnumbers,amsmath,amssymb]{revtex4}
\usepackage{bm}
\begin{document}
\preprint{}
\title{A new insight into the negative-mass paradox of
     gravity and the accelerating universe}
\author{Guang-Jiong Ni}
\affiliation{Department of Physics, Fudan University, Shanghai, 200433, China\\
Department of Physics, Portland State University, Portland, OR 97207 USA}
    \altaffiliation[ ]{E-mail address: pdx01018@pdx.edu}

\begin{abstract}
The discovery of acceleration of the universe expansion in recent astrophysics research
prompts the author to propose that Newton's gravitation law can be generalized to
accommodate the antimatter: While the force between matters (antimatters) is attractive,
the force between matter and antimatter is a repulsive one. A paradox of negative-mass in
gravity versus a basic symmetry ($m\rightarrow -m$) based on quantum mechanics is
discussed in sufficient detail so that the new postulate could be established quite
naturally. Corresponding modification of the theory of general relativity is also suggested.
If we believe in the symmetry of particle and antiparticle as well as the
antigravity between them, it might be possible to consider a new scenario of the
expansion of universe which might provide some new insight into the interpretation of
cosmological phenomena including the accelerating universe observed.
\end{abstract}
 \maketitle

\section{Introduction}
Let us begin with a ``paradox of negative-mass" in the theory of gravity. As is well
known, the Newton's gravitation law reads:
\begin{equation}\label{eq:1}
    F(r)=-G m_1 m_2 /r^2,
\end{equation}
where $m_1$ or $m_2$ is the gravitational mass of a particle or a macroscopic body with
spherical symmetry, $r$ is the distance between them and $G$ is the gravitational
constant. The minus sign in (1) means that the force $F$ between them is always an
attractive one, which in turn implies that the gravitational potential energy of them is
negative:
\begin{equation}\label{eq:2}
  U(r)=-G m_1 m_2 /r.
\end{equation}

On the other hand, all experiments and the theory of special relativity (SR) have been
verifying the equivalence of mass $m$ and energy $E$, i.e., the Einstein's equation:
\begin{equation}\label{eq:3}
  E=mc^2,
\end{equation}
where $m$ is the inertial mass. It is defined by another Newton's law of dynamics:
\begin{equation}\label{eq:4}
  F=m\,a,
\end{equation}
where $a$ is the acceleration of the particle (body).

Consider a system (body) being composed of many particles, the gravitational binding
energy shown by (2) would render the total mass of the body decreasing. Then an acute
problem arises: can this body have a negative mass? If so, a very bizarre phenomenon would
occur as discussed by Bondi, Schiff and Will respectively [1]: Suppose that such a body
(with mass $m_1<0$) is brought close to a normal body (with mass $m_2>0$). According to
Eqs. (1) and (4), the positive-mass body ($m_2$) would attract the negative-mass
body($m_1$) whereas the negative-mass body($m_1$) would repel the positive-mass body
($m_2$). The pair ( a ``gravitational dipole" ) would accelerate itself off, without any
outside help or use of propulsion! The conservation law of momentum and that of energy
would all be violated.

Incredible! No one can believe in that. Hence we may name the above problem a
``negative-mass paradox" in gravity. There are a number of paradoxes in physics, aiming
at pushing the contradiction hidden in the theory to a situation that enables one to
clarify what was wrong or missing in the basic concept.

To get rid of this ``negative-mass paradox", a ``positive-energy theorem" was posed in
the middle of 1960s, saying that the total asymptotically determined mass of any isolated
body in general relativity(GR) must be non-negative. This theorem had been proved since
1979 in a variety of ways and is in total conformity with Einstein's equation (3) because
the observed inertial mass and energy are always positive definite.

However, being a paradox as acute as that hidden in Eq. (1), it does provide a much
stronger hint than a negation of negative mass. Remembering that the appearance of
negative energy is inevitable in relativistic quantum mechanics (RQM) and is
intimately related to the existence of antiparticles, I will manage to claim that the
whole solution to the ``negative-mass paradox" must be a generalization of Newton's
gravitation law into the following form:
\begin{equation}\label{eq:5}
    F(r)=\pm G m_1 m_2 /r^2,
\end{equation}
where the minus sign holds for $m_1$ and $m_2$ (both positive) being both matters or both
antimatters whereas the plus sign holds for one of them being antimatter, which means
that matter and antimatter repel each other.

\section{Symmetry of space-time inversion}
Starting from RQM, we consider the wavefunction (WF) of a freely moving (along $x$ axis)
particle:
\begin{equation}\label{eq:6}
  \psi\sim exp\;[i(p\;x-E\;t)/\hbar],
\end{equation}
where $p$ is the momentum and $E(>0)$ the total energy.

But what is the WF $\psi_c$ of an antiparticle? Before 1956, it was assumed to be a
consequence of the operation of a so-called charge-conjugate transformation C which can bring a
charged particle ( say an electron with charge $-e$) to its antiparticle (say the
positron with charge $e$) [2]:
\begin{equation}\label{eq:7}
  \psi_c=C\psi\sim\psi^*\sim exp\;[i(-p\;x+E\;t)/\hbar].
\end{equation}

We see that the negative-energy $-E<0$ emerges immediately due to the basic operators in
quantum mechanics(QM):
\begin{equation}\label{eq:8}
  \hat{p}=-i\hbar\frac{\partial}{\partial x},\quad \hat{E}=i\hbar\frac{\partial}{\partial
  t}.
\end{equation}

The negative-energy difficulty at the level of QM was remedied to some extent by the
so-called ``hole theory for positron" (which is obviously impossible for the Klein-Gordon
particle) and solved formally at the level of quantum field theory (QFT) by the
redefinition of creation (annihilation) operators.

Since the discovery of parity violation ,i.e., the violation of space-inversion $P
(x\rightarrow -x)$ symmetry in 1956-1957, physicists realize that not only P but also C
transformations are not conserved in the weak-interaction processes. So Eq. (7) is not
applicable in general and the WF of antiparticle should be redefined as :
\begin{equation}\label{eq:9}
  \psi_c=CPT\psi\sim exp\;[-i(p\;x-E\;t)/\hbar],
\end{equation}
where the so-called ``time-reversal transformation" T is defined as:
\begin{equation}\label{eq:10}
  \psi\rightarrow T\psi=\psi^* (x,-t)\sim exp\;[i(-p\;x-E\;t)/\hbar].
\end{equation}
Note that: First, the name ``time-reversal" is actually a misnomer[3, 4]. What the
transformation (10) means is merely a reversal of motion $(p\rightarrow-p)$. Second, the
correctness of definition of antiparticle WF (9) depends on the validity of the CPT theorem
which in turn is ensured by basic principles of SR and QFT. Third, as the
complex-conjugate operations in C and T cancel each other, what a combined CPT
transformation in (9) means is merely a sign change of coordinates $(x,t)$ in comparison
with Eq. (6) [2]. But the original meaning of C, P and T implies that Eq. (9) should
describe an antiparticle with the same $p$ and $E(>0)$ as that of the particle described by
(6). Hence for antiparticles, we should forget the ``hole theory" and use the following
operators instead of (8):
\begin{equation}\label{eq:11}
  \hat{p}\;_c=i\hbar\frac{\partial}{\partial x}\;,\quad \hat{E}_c=-\;i\;\hbar\frac{\partial}{\partial
  t}\;.
\end{equation}
In fact, Eq. (11) had been proven to be the direct and unique outcome of the full solutions to
the EPR paradox and Klein Paradox [5,4,6].

Fourth, once we accept Eqs.(9) and (11), the CPT theorem becomes a new fundamental
postulate, i.e., a basic symmetry which can be stated in the following form:

Under the (newly defined) space-time inversion denoted by $\cal{PT}$, meaning merely
$x\rightarrow-x$, $t\rightarrow-t$, the theory of RQM remains invariant with its concrete
solution, e.g., a particle WF transforming to its antiparticle WF (denoted by $\cal C$)
automatically. It means that our postulate reads:
\begin{equation}\label{eq:12}
  \cal{PT}=\cal C.
\end{equation}
For example, the Schr\"odinger equation is nonrelativistic. But the following coupled
Schr\"odingerlike equation
\begin{equation}\label{eq:13}
  \left\{\begin{array}{ll}
  i\hbar\frac{\displaystyle\partial}{\displaystyle\partial t}\;\varphi=mc^2\varphi-
  \frac{\displaystyle\hbar^2}{\displaystyle {2m}}\;\nabla^2\;(\varphi+\chi),\\[3mm]
  i\hbar\frac{\displaystyle\partial}{\displaystyle\partial t}\;\chi=-mc^2\chi+
  \frac{\displaystyle\hbar^2}{\displaystyle {2m}}\;\nabla^2\;(\chi+\varphi),
  \end{array}\right.
\end{equation}
is just the relativistic Klein-Gordon equation
\begin{equation}\label{eq:14}
  (i\hbar\frac{\displaystyle\partial}{\displaystyle\partial t})^2\;\psi=-\,c^2{\hbar}^2
  \nabla^2\;\psi+m^2c^4\psi,
\end{equation}
with relation first pointed out by Feshbach and Villars in 1958:
\begin{equation}\label{eq:15}
  \left\{\begin{array}{ll}
  \varphi=(\psi+i\;\hbar\;\dot{\psi}/mc^2)/2,\\[1mm]
  \chi=(\psi-i\;\hbar\;\dot{\psi}/mc^2)/2.
  \end{array}\right.
\end{equation}
Now we see that under the space-time inversion $(x\rightarrow-x, t\rightarrow-t)$ and the
transformation:
\begin{equation}\label{eq:16}
  \varphi(-x,-t)\rightarrow\chi(x,t),\quad \chi(-x,-t)\rightarrow \varphi(x,t),
\end{equation}
the Eq. (13) does remain invariant while a particle WF (6) with $|\varphi|>|\chi|$ (due
to $E>0$, see (15)) turning to its antiparticle WF (9) with $|\chi_c|>|\varphi_c|$ (due to
$E<0, E_c=-E>0$, see (16)).

\section{Symmetry of mass inversion}

Alternatively, we can restate the above basic symmetry in the following way: Under the
mass inversion:
\begin{equation}\label{eq:17}
  m\rightarrow-m,\quad \varphi(x,t)\rightarrow \chi(x,t),\quad
  \chi(x,t)\rightarrow\varphi(x,t),
\end{equation}
the theory, e.g., Eq. (13), remains invariant. Although transformation (16) is equivalent
to transformation (17), they share different advantages. The former is relevant to
unobservable coordinates $(x,t)$ and so is more essential in RQM and equivalent to even
more abstract symmetry of $i$ versus $-i$ (see Eq.(6) versus (9)), while the latter is
relevant to observable mass $m$ and so is easily to be generalized to the case of
classical theory.

For example, as is well known, the motion equation for an electron is given by the
Lorentz-force law:
\begin{equation}\label{eq:18}
  \bm F=m\bm a=-e\;(\bm E+{\bm v}\times\bm B/c).
\end{equation}
Before 1956, its counterpart for positron is obtained via a C transformation
$(-e\rightarrow e)$ , yielding:
\begin{equation}\label{eq:19}
  \bm F_c=m\bm a_c=e\;(\bm E+\bm v\times\bm B/c),
\end{equation}
which is still allowed since C is conserved in the electromagnetic interaction. However,
we'd better use the following working rule: to deal with particle (matter) and
antiparticle(antimatter) on an equal footing, a classical theory must be invariant under
the mass-inversion transformation $m\rightarrow -m$. Using this rule to Eq. (18), we get
(19) immediately. Note that: First, $m$ is always positive. Second, being the external
field, the electric-(magnetic) field strength $\bm E(\bm B)$ undergoes no change in the
transformation. Third, in RQM like (13) or (14), the motion equation for antiparticle is
the same one as that for particle. This is because each particle state like (6) contains
its hidden antiparticle field under the condition $|\varphi|>|\chi|$ whereas an
antiparticle state like (9) contains its hidden particle field under the condition
$|\chi_c|>|\varphi_c|$. By contrast, in classical theory, the equation for particle is
separated from that for antiparticle as shown by (18) versus (19). We just complement
(18) with (19) so that the theory becomes invariant under the mass inversion
transformation. In other words, a new equation must be added before it can be complete in
the sense of treating the particle and antiparticle equally.

Fourth, to clarify further why we prefer the new postulate (12) instead of C
transformation and CPT theorem, we wish to emphasize an important difference between a
postulate (law) and a theorem. All quantities in a theorem must be defined in advance
separately and unambiguously and the outcome of the theorem is actually contained in its
premise implicitly. For example, the definitions of C, P and T are all clear in
mathematics and the validity of CPT theorem is ensured by the basic principles of SR and
QFT. Once C, P and T are not conserved in experiments, they cease to be meaningful as
physical transformations. In this situation, the CPT theorem immediately exhibits itself as a
new postulate (12) in which the definition of transformation of particle to antiparticle
is just contained in the same equation. In general, a postulate or law can often (not
always) accommodate a definition of physical quantity, and the validity of the postulate
(law) together with the definition must be verified by experiments. Hence the
establishment of a law is a process ``from particular to general" or an outcome of
``analysis and induction method". By contrast, to prove a theorem from well-established
theories is a process ``from general to particular" or a consequence of ``deduction
method".

For example, the definition of gravitational mass $m$ is contained in the gravitational
law, Eq. (1). The definition of inertial mass $m$ is contained in Newton's law , Eq.
(4). The definition of electric-magnetic field strength, $\bm E$ and $\bm B$, is
contained in the Lorentz force law, Eq. (18). What we have done is a similar thing---the
definition of particle-antiparticle transformation $\cal C$ is contained naturally in a new
postulate (12) ---- not one that comes from elsewhere.
\section{Generalization of Newton's gravitation law to antimatter and antigravity}
We are now in a position to establish Eq. (5), performing the rule of mass-inversion on
Eq. (1) together with Eq. (4). Note that, however, both $m_1$ and $m_2$ belong to a
closed system under consideration. When one of them, say $m_1$, is an antimatter, we just
perform a transformation $m_1\rightarrow-m_1$ once, regardless of whether it is a source
($m_1$ appears only at one side of equation) or a detected body ($m_1$ appears at
both sides of equation). This simple procedure brings Eq. (1) to Eq. (5) with a plus sign,
implying that $m_1$ and $m_2$ repel each other. Furthermore, if $m_2$ is also an
antimatter, one more transformation $m_2\rightarrow-m_2$ recovers the minus sign,
implying that two antimatters attract each other. Now the whole Eq. (5) is invariant with
respect to transformation $m_i\rightarrow-m_i (i=1, 2)$.

The reason we can get rid of the ``negative-energy paradox" is as follows. If one
thought that the detected body $m_1$ is really a negative mass ($m_1<0$ while $m_2>0$) seriously,
the paradox was inevitable. But here every mass is positive regardless of whether it is matter or
antimatter. Even if we make a careless mistake of performing transformation
$m_1\rightarrow-m_1$ twice, it is still not a serious mistake---we merely do a trivial
thing to return back to the minus sign. Yet by another transformation on either $m_1$ or
$m_2$ will the other equation with plus sign be supplemented. Eventually, the correct
interpretation of whole Eq. (5) is independent of the procedures used to reach it. The bizarre
phenomenon which occurred in the paradox will never happen again.

In some sense, the paradox stemmed from the incorrect conception that the difference
between positive and negative masses is absolute, whereas now we regard the difference
between $m$ and $-m$ merely relative. We merely perform a symmetry transformation
$m\rightarrow-m$ to show the equal existence of matter versus antimatter, but eventually
there is no negative mass at all. This reminds us of the experience (i.e., lesson) of
Einstein. He struggled for eight years before inventing the theory of GR. What Einstein
eventually realized is that space and time have no absolute meaning but systems
of relations (as stressed by Smolin [7]).
\section{Generalization of Einstein field equation in general relativity}
Consider a positronium and an atom of matter. If Eq.~(5) is correct, there will be no
gravitational force between them. This means that the gravitational mass $m$(grav.) of
positronium is zero! However, the energy or the relevant inertial mass $m$(inert.) (see
Eq.~(3)) of positronium is definitely nonzero. Hence we see that in the case of
coexistence of particles and antiparticles, the equivalence principle (EP) in the (weak)
sense that [8]
\begin{equation}\label{eq:20}
    m\,({\rm grav.})=m\,({\rm inert.})
\end{equation}
cannot be valid.

As is well known, the EP served as a starting point in establishing the theory of general
relativity(GR). The possible invalidity of EP in the presence of antimatter implies that
GR is dealing with the gravitation of pure matter without the coexistence of antimatter.
Indeed, let us look at the Einstein field equation [9]:
\begin{equation}\label{eq:21}
    R_{\mu\nu}-\frac{1}{2}g_{\mu\nu}R=-8\pi GT_{\mu\nu}
\end{equation}
On the left side, the Ricci tensor $R_{\mu\nu}$, curvature scalar $R$ and the metric
tensor $g_{\mu\nu}$ are all functions of coordinates $x_\mu$. While on the right side,
the energy-momentum tensor $T_{\mu\nu}$ is introduced to describe the existence of matter
in the vicinity (a marcroscopic small volume) of $x_\mu$. Then under a transformation of
mass inversion $m\rightarrow-m$ to reflect that of matter to antimatter, $T_{\mu\nu}$
should change its sign due to its proportionality to the mass $m$. Hence Eq.~(21) changes
sign on the right side whereas not on the left side. This reflects the fact that GR is a
classical field theory and so cannot treat the matter and antimatter on an equal footing.

To keep Eq.~(21) invariant under the mass inversion, we manage to modify its right side
by a generalization as:
\begin{equation}\label{eq;22}
   T_{\mu\nu}\rightarrow T_{\mu\nu}^{eff}=T_{\mu\nu}-T_{\mu\nu}^c,
\end{equation}
(where the superscript c means antimatter,) since under the mass inversion,
$T_{\mu\nu}\rightarrow -T_{\mu\nu}^c$ and $T_{\mu\nu}^c\rightarrow - T_{\mu\nu}$. Notice
that the form of the energy-momentum tensor is the same for both matter and antimatter. We
stress once again that the distinction between $m$ and $-m$ is merely relative, not
absolute.

\section{Antigravity and cosmology}

Evidently, Eq.~(5) cannot be tested by experiments on earth. We can only await and see
new developments in astrophysical research. Surprisingly, in recent years, the careful
observation of some distant Type Ia supernovae with redshift $z=0.39-0.9$ reveals that
the expansion of the universe is accelerating rather than decelerating as physicists
pondered several years ago (see the excellent papers [10,11,12]). Many physicists tend to
explain the acceleration by resorting to the cosmological constant (with dimension
(length)$^{-2}$) originally introduced by Einstein, unsuccessfully trying to interpret it
as a type of ``vacuum zero-point energy" in QFT. It is often called ``dark energy", which
may account for up to $70\%$ of the entire matter and energy in the universe. Actually,
nobody understands what the dark energy really is.

Although I have no research experience in the field of GR and astrophysics, I do share
the opinion of Davies [13] that "if we want to keep inflation and account for today's
accelerating expansion, we need a theory that explains why antigravity was once tense,
then dropped and hovered at just above zero,.... One possibility is that the force fades
with time. Another is that it varies in space, so that far beyond the limit of our
telescopes it may be much bigger,.... What we need is a theory that derives the strength
of the antigravity force as part of a unified description of all the forces of nature." I
am trying to pose some conjecture as follows.

(a). During the inflation era right after the big bang, once particles and antiparticles
were created, they began to repel each other according to Eq.~(5). I don't believe the
interpretation that the present baryon asymmetry stemmed from the small CP violation
(which is equivalent to T violation, as discussed above, it has nothing to do with the
basic symmetry in time or that in particle and antiparticle). I prefer to think that the
total number of particles is equal to that of antiparticles in the entire expanding
universe. But antiparticles had a head start during the inflationary expansion and flew
away faster than particles did. The latter lagged behind to some extent and our galaxies
gradually evolved out of them. On the other hand, some distant stellar objects (quasars?)
might be evolved from antiparticles and some of them may be just those observed Type Ia
supernovae undergoing acceleration caused by the repulsive force exerted by inner
galaxies composed mainly of matters.

(b). Based on the above picture and solving Eq.~(21) with modification (22) in the
Friedmann model as discussed in [9], we can see that in a large part of universe where
the densities of matter and antimatter are nearly equal, the acceleration of the
expansion tends to zero. In other words, the flatness of our universe as shown by the
recent careful measurement of the fluctuation of cosmic microwave background
radiation (see [14]) is critically depending on two factors: first, the inflationary
expansion triggered by the repulsive force between matter and antimatter; second, the
nearly equal densities of matter and antimatter in a large part of universe after the
inflationary expansion. So a careful measurement of the variation of the Hubble constant [11]
with distance may reveal the subtle difference in different regions.

(c). Among all particles, neutrinos are of particular interest [15]. After the big bang,
superluminal neutrinos and antineutrinos are equally created in three flavors and
distributed everywhere isotropically. Their number density may be $10^8$ times that of
protons. However, because the forces between neutrinos and matter just cancel that
between antineutrinos and matter, all neutrinos and antineutrinos (even they all have
positive tachyon mass, see VII,(d)) scape from the analysis of so-called dark matter.

(d). As the next conjecture, I guess that in the intermediate region where matter and
antimatter are overlapping, there may be some probability of collisions between matter and antimatter. Because of the long range repulsive force between them,
these collisions may be grazing and so the annihilation process occurs at
short distance would have some special character. Is it possible to relate the above
conjecture with the mechanism of gamma-ray-bursts(GRBs)? It has been noticed for years that GRBs
are distributed at a remote distance on the cosmological scale and they tend to have a
roughly constant explosive energy (see [16]and references therein). So the careful study
of the GRB-Hubble diagram as conducted by Schaefer[16] is very important because the spatial
distribution of GRB may provide further evidence of nonuniform distribution of matter and
antimatter in the universe.

(e). Furthermore, I think that the properties of cosmic matter composed of pure matter and
that of a mixture of matter with antimatter are certainly different. The latter bears
some resemblance to plasma (not quite the same) with a ``screening length". The
fluctuations in plasma with length scale larger than the screening length should be
strongly suppressed. In fact, the star formation rate is rising steadily from remote
distance with redshift $z\sim 2$ to $z\sim 6$ [16] which seems to me an reflection that
the domination of antimatter is increasing steadily there.

\section{Summary and discussion}

(a). I was pleasantly surprised to read the wonderful paper by Sch\"odel et al [17].
Their delicate observation over ten years and convincing analysis verified the existence
of a huge black hole at the center of the Milky way. This should be viewed as a triumph
of experimental science, also a triumph of the theory of GR , which was first verified by
its successful explanation on the precession of Mercury's perihelion. However, despite
all of these, GR is still a classical theory dealing with the gravitation of pure matter
as shown by its noninvariance under the transformation of mass inversion.

(b). Based on the rule of mass-inversion invariance, the Newton's gravitation law is
generalized to Eq.~(5) and the Einstein field equation in GR, Eq.~(21), is also modified
by (22) such that they can treat matter and antimatter equally and explicitly. The
antigravity between matter and antimatter may be important to account for many
cosmological phenomena including the acceleration of expansion of the universe.

(c). Eq.~(5) should be derived from Eq.~(21) with (22) in a weak-field approximation. We
follow the standard procedure as discussed in [9] and notice that in the case of one
of the bodies being antimatter (the other matter), an extra
minus sign will appear in front of the product of their energy-momentum tensors due to
Eq.~(22). This means antigravity in Eq.~(5).

(d). We are dealing with three kinds of symmetries, the particle-antiparticle symmetry,
the space-time inversion symmetry, and the mass inversion symmetry. They are essentially equivalent to
each other. This can also be seen from the two Casimir invariants in the
Poincar\'e group, which is a group generated by the inhomogeneous Lorentz transformation
[18]:
\begin{equation}\label{eq;23}
      x^\mu=\Lambda_\nu^\mu x^\nu+a^\mu
\end{equation}
Usually, the conditions: det$\,|\Lambda_\nu^\mu|=1$ and $\Lambda_0^0\geq 1$ are imposed.
So one is dealing with a proper and orthochronous Poincar\'e transformation. There are
two Casimir invariants [commuting with all ten generators $M_{\mu\nu}$ (the six
generators of the Lorentz group) and $P_{\mu\nu}$ (the four generators of the space-time
translation group)] being: $P_{\mu\nu}P^{\mu\nu}$ with eigenvalues $m^2$, ($m$ is the
particle mass) and $W_\sigma W^\sigma$
($W_\sigma=-\frac{1}{2}\epsilon_{\mu\nu\rho\sigma}M^{\mu\nu}P^\rho$, the Pauli-Lubanski
operator) with eigenvalues $-m^2s(s+1)$, ($s$ is the particle spin). Now we know that the
physical essence of the Lorentz group is that of SR. The latter can be derived from any
one of the three discrete symmetries mentioned above based on QM. Hence we can generalize
the Poincar\'e group beyond the proper and orthochronous one to incorporate the
antiparticle explicitly, i.e., we may perform the inversion of $P_\mu$ and $M^{\mu\nu}$
so that $P_\mu\rightarrow -P_\mu$, $W_\sigma\rightarrow -W_\sigma$ which means the
transformation of a particle to its antiparticle. The invariance of $P_\mu P^\mu$ and
$W_\sigma W^\sigma$ implies that their eigenvalues remain invariant under the
corresponding inversion of mass: $m\rightarrow -m$. In other words, either the Lorentz
group or the Poincar\'e group is relativistic in essence which is nothing but the equal
existence of particle and antiparticle. Moreover, a negative eigenvalue of $P_\mu
P^\mu=m^2<0$ is also allowed. This implies an analytical continuation of mass:
$m\rightarrow im_s$ or $-im_s$ with $m_s$ real (the tachyon mass) to describe a
superluminal neutrino or antineutrino [15]. Here the symmetry between i and -i reflects
even more deeply the symmetry of mass inversion which is really a relative one rather
than an absolute one.

(e). The ``photon" with $m^2=0$ in the representation of Poincar\'e group is by no means
an ordinary particle. Usually, a photon with linear polarization ($\gamma_x$ or
$\gamma_y$) can be viewed as a linear superposition of a photon with left-handed
polarization ($\gamma_{_L}$) and that with right-handed polarization ($\gamma_{_R}$). But
in our unified theoretical scheme of space-time inversion, $\gamma_{_L}$ and
$\gamma_{_R}$ could be viewed as the ``particle" and ``antiparticle" (or vice versa). So
I guess that a photon has no gravitational mass. Alternatively, we may say that a photon
exerts no gravitational force on other matter (antimatter). It just moves along the
geodesic line in the space-time which is determined by all matter and antimatter in the
entire universe. By the way, according to our understanding of so-called ``wave-particle
duality"[4], a ``photon" should be treated theoretically as an electromagnetic wave in
propagation until detected then it appears as a "particle". To speak of a photon's
position before it is detected is meaningless.

(f). The possible violation of the equivalence principle in the case of coexistence of matter
and antimatter has far-reaching implication. The energy or inertial mass
[linked by Eq.~(3)] obeys the conservation law and thus is observable. By contrast,
the gravitational mass is not conserved and thus is not observable in the strict sense. It
merely obtains its definition from Newton's gravitation law, which, of course, is a
classical and approximate one.

(g). I dare not discuss the quantization of GR because I have little knowledge
about the theory of quantum gravity (see the excellent introduction by Smolin [7]).
However, according to our understanding about SR  based on QM [4, 6], a possible ``fourth
road" to quantize GR might be a localization of the space-time inversion symmetry mentioned
above (just like GR being the localization of the Lorentz symmetry in SR) so that both
matter and antimatter can be treated explicitly and equally. Regrettably, I have been
pondering this problem for many years with nearly no progress. So I hope that more
physicists will join this difficult but interesting pursuit.

\begin{acknowledgments}
I thank S. Q. Chen, P. Leung, D. Lu and Z. Q. Shi for helpful discussions and bringing
relevant references to my attention. I am greatly indebted to three referees for their
kind comments leading to much improvement on this paper.
\end{acknowledgments}


\begin{thebibliography}{18}
\bibitem{1}Bondi, H. Negative mass in general relativity, Rev. Mod. Phys. 29, 423-428
(1957); Schiff, L. I. Sign of the gravitational mass of a positron, Phys. Rev. Lett. 1,
254- 255(1958); Will, C., The renaissance of general relativity, in The new Physics, Ch.2
(Edited by Davies, P., Cambridge Univ. Press, 1989)p.31.
\bibitem{2}Bjorken, J. D. and Drell, S. D., Relativistic Quantum Mechanics (Mcgraw-Hill Book Company, 1964).
\bibitem{3}Sakurai, J. J., Modern Quantum Mechanics (Revised Ed.)(John Wiley \& Sons, Inc.1994).
\bibitem{4}Ni, G. J. and Chen, S. Q., Advanced Quantum Mechanics (Chinese Ed., Press of Fudan University, 2000),
(English Ed., Rinton Press, 2002).
\bibitem{5}Ni, G. J., Guan, H., Zhou, W. M. and Yan, J., Antiparticle in the light of
Einstein-Podolsky-Rosen paradox and Klien paradox, Chin. Phys. Lett. 17, 393-395(2000),
quant-ph/0001016.
\bibitem{6}Ni, G. J., Ten arguments for the essence of special relativity, in
Proceedings of the 23rd Workshop on High-energy Physics and Field Theory, (Protvino,
Russia, June 2000, Edit: Filimonova, I. V. and Petrov, V. A.) 275-292; hep-th/0206250. An
modified version will appear in Progress of Physics (Nanjing, China).
\bibitem{7}Smolin, L., Three Roads to Quantum Gravity (Basic Books, 2001) p.149.
\bibitem{8}Will, C. W., Theory and experiment in gravitational physics, Revised Edition
(Cambridge Univ. Press, 1993).
\bibitem{9}Weinberg, S., Gravitation and cosmology (John Wiley, 1972).
\bibitem{10}Bahcall, N. A., Ostriker, J. P., Perimutter, S. and Steinhardt, P. J., The cosmic triangle: revealing the
state of the universe, Science, 284, 28 May, 1481-1488(1999).
\bibitem{11}Freedman, W., The Hubble constant and the expanding universe, American Scientist, 91, Jan-Fe. 36-43(2003).
\bibitem{12}Livio, M., Moving right along, Astronomy, July, 34-39(2002).
\bibitem{13}Davies, P., Seven wonders, New Scientist, 21 Sept. 28-33(2002).
\bibitem{14}New Scientist, 2003, 15 Feb., 12-13;2003, 5 April, 22-23.
\bibitem{15}Ni, G. J., A minimal three-flavor model for neutrino oscillation based on
superluminal property, preprint, hep-ph/0306028.
\bibitem{16}Schaefer, B. E., Gamma-ray burst Hubble diagram to z=4.5, The Astrophysical
journal, 583: L67-L70(2003).
\bibitem{17}Sch\"odel, R., et al. A star in a 15.2-year orbit around the
supermassive black hole at the centre of the Milky Way, Nature, 419, 17 Oct. 694-696
(2002).
\bibitem{18}Marshak, R. E., Conceptual foundations of modern particle physics
(World Scientific,1993).


\end{thebibliography}
\end{document}